\documentclass[aps,showpacs,pra,twocolumn,superscriptaddress,10pt]{revtex4-1}

\usepackage{graphicx}
\usepackage{dcolumn}
\usepackage{bm}
\usepackage{hyperref}
\usepackage{textcomp}

\begin{document}


\title{Photonic Band Gaps in 3D Network Structures with Short-range Order}
\author{Seng Fatt Liew}
\email{sengfatt.liew@yale.edu}
\affiliation{Department of Applied Physics, Yale University, New Haven, CT 06511, USA}
\author{Jin-Kyu Yang}
\affiliation{Department of Applied Physics, Yale University, New Haven, CT 06511, USA}
\affiliation{Department of Optical Engineering, Kongju National University, Kongju 314-701, Republic of Korea.}
\author{Heeso Noh}
\affiliation{Department of Applied Physics, Yale University, New Haven, CT 06511, USA}
\author{Carl F. Schreck}
\affiliation{Department of Physics, Yale University, New Haven, CT 06511, USA}
\author{Eric R. Dufresne}
\affiliation{Department of Mechanical Engineering and Materials Science, Yale University,  New Haven, CT 06511, USA}
\affiliation{Department of Physics, Yale University, New Haven, CT 06511, USA}
\affiliation{Department of Chemical and Environmental Engineering, Yale University, New Haven, Connecticut 06511, USA}
\affiliation{Department of Cell Biology, Yale University, New Haven, Connecticut 06511, USA}
\author{Corey S. O'Hern}
\affiliation{Department of Mechanical Engineering and Materials Science, Yale University,  New Haven, CT 06511, USA}
\affiliation{Department of Physics, Yale University, New Haven, CT 06511, USA}
\author{Hui Cao}  
\affiliation{Department of Applied Physics, Yale University, New Haven, CT 06511, USA}
\affiliation{Department of Physics, Yale University, New Haven, CT 06511, USA}



\date{\today}

\begin{abstract}
We present a systematic study of photonic band gaps (PBGs) in three-dimensional (3D) photonic amorphous structures (PAS) with short-range order.  From calculations of the density of optical states (DOS) for PAS with different topologies, we find that tetrahedrally connected dielectric networks produce the largest isotropic PBGs.  Local uniformity and tetrahedral order are essential to the formation of PBGs in PAS, in addition to short-range geometric order.  This work demonstrates that it is possible to create broad, isotropic PBGs for vector light fields in 3D PAS without long-range order.  
\end{abstract}
\maketitle

\section{introduction\label{introduction}}

A photonic band gap (PBG) describes a frequency range within which light propagation is prohibited due to depletion of optical states.  The most known structures having PBGs are photonic crystals (PhCs) with periodic modulations of dielectric constant \cite{joannopoulos}.  Since PhCs are anisotropic, PBGs vary with directions.  To have a 
complete PBG, the gaps in all directions must overlap in frequency.  This condition is difficult to achieve for many PhCs, for example, simple cubic lattices.  It is therefore easier to produce complete PBGs in more isotropic structures, e.g. photonic quasicrystals that possess higher rotational symmetry (but no translational symmetry) \cite{steinhardt2,chan2}.  Photonic amorphous structures (PAS) are most isotropic, due to the absence of long-range translational or rotational order.  Recent studies demonstrate that PBGs can be formed in two-dimensional (2D) and three-dimensional (3D) PAS with short-range order \cite{zhang,shinya,notomi1,lederer,steinhardt,notomi2,segev}.  However, the exact physical mechanism or condition for the PBG formation in PAS is not well understood.  An improved fundamental understanding of PBG formation would allow 
researchers to design photonic amorphous materials with optimized and tunable PBGs. \\

\indent In addition to geometric order, structural topology plays an important role in forming a PBG.  For the composite dielectric materials consisting of two components with different refractive indices, there are two cases regarding the topology of the high-index component.  (i) Cermet topology: the high-index material consists of isolated inclusions, each of which is completely surrounded by the low-index material.  (ii) Network topology: the high-index material is connected and forms a continuous network running through the whole composite.  Previous studies of periodic structures have indicated that the cermet topology is more favorable for the PBG formation of a scalar wave, while the network topology for a vector field \cite{sigalas}.  Such conclusions also apply to PAS.  For example, in 2D PAS, PBGs for the transverse magnetic (TM) polarization (electric field out of plane) are easily obtained with isolated islands of high-index materials, because the electric field has same polarization direction everywhere and can be regarded as a scalar wave.  For the transverse electric (TE) polarization (electric field in plane), the electric field has varying polarization direction and behaves like a vector field, thus it is easier to produce PBGs in connected dielectric networks \cite{joannopoulos2}.  It has been proposed that a hybrid structure with a mixture of both topologies can possess a full PBG for both TE and TM polarizations~\cite{steinhardt}.\\

\indent It is much more difficult to form complete PBGs in 3D structures.  Substantial reductions in the density of optical states (DOS) have been demonstrated in PAS composed of randomly packed dielectric spheres of uniform size~\cite{lederer}, as a result of evanescent coupling of the Mie resonances of individual spheres.  Dielectric network structures, for example, the photonic amorphous diamond (PAD), exhibit much stronger depletion of the DOS \cite{notomi1,notomi2}.  It was conjectured that the tetrahedral bonding configuration in the PAD plays an important role in the formation of isotropic PBG.  However, the PAD is constructed from a ``continuous-random-network'' (CRN) originally developed for modeling of amorphous Si or Ge \cite{barkema}, thus it is difficult to separate the relative contributions of tetrahedral bonding and local geometric order to the PBG formation.  Identifying the key parameters that determine when a PBG will form in PAS is important not only for developing novel photonic glasses~\cite{lopez}, but also for understanding color generation in nature~\cite{prum}.  Both cermet and network topologies have been found in color-producing PAS of many animal species~\cite{dufresne,noh}.  It is also conjectured that pseudo PBGs may be formed and responsible for non-iridescent coloration of many PAS~\cite{jianzi}. \\

\indent In this article, we present a detailed numerical study of the DOS and PBGs in 3D PAS. We vary the topology, short-range geometric order, refractive index contrast, and filling fraction to maximize the depletion of DOS and the strength of PBG in the absence of long-range structural order.  This study allows us to identify the essential elements for the formation of PBGs in PAS.  We organize the manuscript as follows.  In Sec. \ref{ii}, we describe the methods used to generate PAS numerically and analyze their structural and topological properties.  The calculated DOS for both cermet and network topologies are presented in Sec. \ref{DOS}, together with  interpretations of the results.  In Sec. \ref{order}, we explore dielectric networks with different degrees of structural order to maximize the reduction in the DOS.  We then conclude in Sec. \ref{last}. 

\section{Structure generation and characterization}\label{ii}
\subsection{Spheres packings}\label{sphere}
We first study dielectric composites with the cermet topology---high-index dielectric spheres embedded within a low-index host material (air).  We employ a two-stage numerical protocol to generate `just-touching', jammed sphere packings in a cubic simulation cell with varying positional order~\cite{corey,corey2}.  First, liquid states of monodisperse spheres are cooled at fixed packing fraction $\phi=0.60$ from an initial high temperature $T_0$ to zero temperature at different rates.  In the second step, each zero-temperature configuration is compressed in steps of $\Delta\phi=10^{-3}$ followed by minimization of the total energy until a static packing with infinitesimal particle overlaps is obtained.  By varying the cooling rate, we are able to create static packings with a range of positional order and packing fractions from random close packing at $\phi=0.64$ to the face centered cubic (FCC) structure at $\phi=0.74$.  In general, the slowly cooled samples can be compressed to higher packing fractions.  Figure \ref{fig1} (a) shows a cluster of $50$ spheres from the interior of a jammed sphere packing containing $1000$ spheres at $\phi = 0.64$.  For comparison, we generate completely disordered configurations by placing spheres randomly in the cubic box with no overlaps at $\phi=0.35$. \\ 

\subsection{Dielectric network}\label{network}
\begin{figure}[htbp]
\centering
\includegraphics[scale = 0.2]{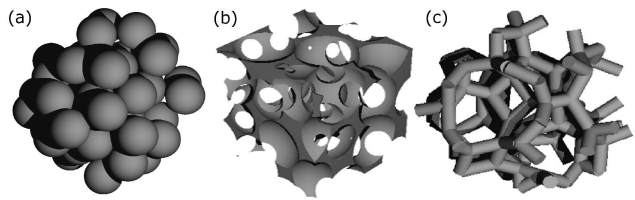}
\caption{Three examples of photonic amorphous structures: (a) jammed packing of dielectric spheres at $\phi=0.64$, (b) inverse structure of (a) with air fraction $\gamma=0.8$, and (c) tetrahedral network of dielectric rods with $\gamma=0.8$ obtained from the Delaunay tessellation of (a).}
\label{fig1}
\end{figure}

We also generate structures with network topologies, where the high-index dielectric material forms the continuous network, using two methods.  For the first method, we invert the cermet structure of jammed dielectric spheres in air. The inverse structure consists of low-index (air) spherical inclusions in a continuous high-index dielectric network.  By adjusting the radius $R$ of the spheres (but fixing their positions), we can vary the air fraction $\gamma$ in the inverse structure. An inverse structure with $\gamma = 0.8$ is shown in Fig.~\ref{fig1} (b).  At this $\gamma$, adjacent air spheres begin to overlap and the dielectric material exhibits an irregular topology. \\ 

\indent The second method, which is based on an algorithm described in Ref.~\cite{steinhardt,hougaard}, produces more uniform network topologies than those from the first method.  In this method, a 3D Delaunay tessellation is performed on the sphere centers from the cermet structures in Sec.~\ref{sphere}.  Each tetrahedron of the tessellation has four facets shared with four neighbors.  We then calculate the center of mass of each tetrahedron, and connect the centers of mass of nearest neighbors by a dielectric rod.  This creates a tetrahedrally connected dielectric network, where each junction (vertex) has four dielectric bonds. 
All dielectric rods have same radius $W$, but different lengths $d$. 
By changing $W$, we can vary the air fraction $\gamma$.  A tetrahedral network with $\gamma = 0.8$ is shown in Fig.\ref{fig1} (c).   

\subsection{Structural characterization}
We now calculate the density autocorrelation function and spatial Fourier spectra of the cermet and network structures described above.  Since the dielectric spheres embedded in air and the corresponding inverted structure possess identical geometrical properties, we focus only on the air spheres and tetrahedral network structures below.\\

\begin{figure}[htbp]
\centering
\includegraphics[scale = 0.3]{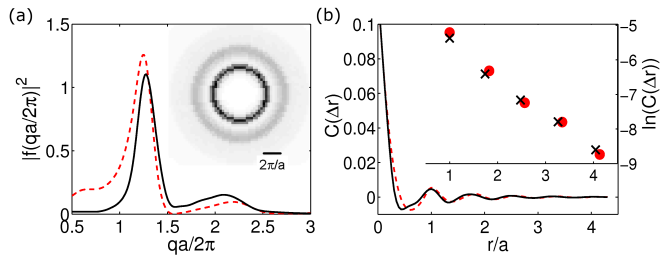}
\caption{Structural characterization of photonic amorphous structures. (a) Angle-averaged power spectra of the spatially Fourier transformed density for jammed sphere packings (dashed line) and tetrahedral networks (solid line) versus $qa/2\pi$, where $q$ is the spatial frequency and $a$ is the mean spacing between spheres. The inset shows a cross-section of the 3D power spectrum for the tetrahedral network. (b) Angle-averaged density autocorrelation for the sphere packing and network structures. The inset shows the amplitudes of the oscillatory peaks of $C(\Delta r)$ for sphere packings (circles) and tetrahedral networks (crosses).}
\label{fig2}
\end{figure}

\indent As shown in the inset to Fig.~\ref{fig2} (a), the 3D spatial Fourier transform of the tetrahedral network structures displays concentric spherical shells without discrete Bragg peaks, which reflects structural isotropy and a lack of long-range order. 
 The radii of the shells provides the characteristic spatial modulation frequencies of the structures.  Similar results are obtained for the tetrahedral networks generated from the jammed sphere packings.  The angle-averaged power spectra for both sphere and network structures are plotted in Fig.~\ref{fig2} (a).  The main peak represents the dominant spatial frequency, and its width is inversely proportional to the average size of ordered domains~\cite{jin}.  The sphere and network structures have similar peak widths, and thus comparable domain sizes. \\

\indent We also calculated the real-space density autocorrelation function $C(\Delta r)$ averaged over all angles for the sphere and network structures~\cite{jin}.  As shown in Fig.~\ref{fig2} (b), both structures display highly damped oscillations of $C(\Delta r)$.  The first peak away from $\Delta r = 0$ is located at the average spacing $a$ between nearest neighbors.  We find that the amplitudes of the oscillatory peaks decay exponentially [inset to Fig. \ref{fig2} (b)] with a decay length (excluding the first peak) $\xi_r \approx 0.9a$ for the sphere packings and $1.1a$ for the tetrahedral networks.  Hence, there are weak spatial correlations and short-range order in these PAS.\\

\section{DOS of PAS with cermet and network topologies}\label{DOS}
In this section, we describe calculations of the DOS for jammed dielectric spheres in air, the inverse structure, and the tetrahedral networks of dielectric rods using the order-$N$ method \cite{chan}.  We choose a cubic supercell with size $8.7a$ containing $1000$ spheres and refractive indices $n=3.6$ and $1$ for the high- and low-index materials, respectively.  We find that the optimal air fraction that yields the largest reduction of the DOS is $\gamma=0.75$ for the dielectric sphere packings and $0.80$ for both the inverse structure and tetrahedral network.  The DOS was ensemble-averaged over five distinct configurations at the optimal $\gamma$ for each topology, and then normalized by the DOS of a ``homogeneous'' medium with the same $\gamma$.  The latter structure is generated by placing cubic dielectric voxels (with lateral dimension $0.043a$, which is much smaller than the wavelength of light $\lambda$) randomly in the supercell.\\
\begin{figure}[htbp]
\centering
\includegraphics[scale = 0.5]{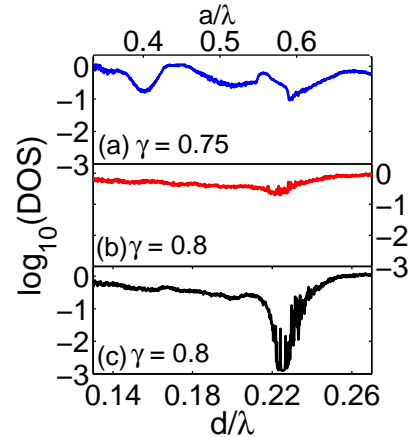}
\caption{DOS for (a) jammed dielectric spheres in air with $\gamma=0.75$, (b) inverted structures with $\gamma = 0.8$, and (c) tetrahedral networks with $\gamma = 0.8$. The wavelength $\lambda$ is normalized by the mean spacing between spheres $a$ (average bond length $d$) on the 
top (bottom) scale.}
\label{fig3}
\end{figure}

\indent As shown in Fig. \ref{fig3}, the maximal DOS reduction occurs in the tetrahedral network structure, which is two orders of magnitude larger than that for the dielectric spheres and inverse structures.  For the tetrahedral networks, the PBG is formed at normalized frequency $d/\lambda  \approx 0.22$, where $d$ is the average length of dielectric rods and $d/a = 0.39$.  The width of the PBG normalized by the gap center frequency is $\sim 5.5\%$.  The modest reduction in the DOS at $a/\lambda \approx 0.41$ for the dielectric spheres stems from Mie resonances of individual spheres~\cite{lederer}.  The uniformity of the dielectric spheres allows the coupling of their Mie resonances, which of the lowest order for isolated dielectric spheres in air occurs at $a/\lambda \approx 0.41$.  In contrast, the air sphere structures have only a small reduction of the DOS in the frequency range where the tetrahedral networks show a pronounced PBG, despite the fact that both structures have dielectric network topology and similar degree of spatial correlation.  It is clear that the dramatic difference in the DOS cannot be explained by the small differences in spatial correlations. \\

\indent Our studies of jammed dielectric sphere packings show that uniformity in the size of dielectric spheres leads to strong coupling of Mie resonances that result in a depletion of the DOS.  In the inverse structure of air spheres, the basic scattering unit is the dielectric filling between air spheres.  For the tetrahedral network structure, the basic scattering unit is centered at each junction where four dielectric rods meet.  Note that in the network topology, the adjacent scattering units are connected, in contrast with the cermet topology.  To compare the uniformity of local scattering units in dielectric networks, we calculate the average refractive index near the center of each unit.  For the tetrahedral network structure, we calculate the mean refractive index $\bar{n}$ within a sphere of radius $r$ whose center coincides with the center of each junction.  We then compute the average $\langle \bar{n}(r) \rangle$ and its variance $V(r)$ over all junctions.  For the air spheres, the dielectric junction center is set at the center of refractive index distribution within each tetrahedron obtained from the 3D Delaunay tessellation of the sphere centers.  
Similarly, we calculate the mean refractive index $\bar{n}$ around each junction center, $\langle \bar{n}(r) \rangle$, and $V(r)$ averaged over all junctions. \\ 

\begin{figure}[htbp]
\centering
\includegraphics[scale = 0.3]{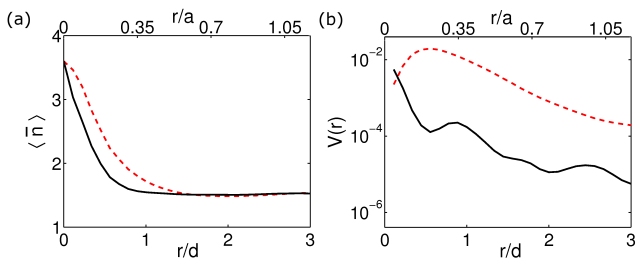}
\caption{ Uniformity of the local scattering environment for the dielectric networks of tetrahedral bonding (solid line) and air spheres (dashed line). (a) Mean index of refraction $\langle\bar{n}(r)\rangle$ and (b) variance $V(r)$ within a distance $r$ from the dielectric junction center. $r$ is normalized by the mean spacing of spheres $a$ (average bond length $d$) on the 
top (bottom) scales.}
\label{fig4}
\end{figure}

\indent In Fig.~\ref{fig4} (a) we show that on average the tetrahedral network and air spheres structure have similar distributions of the mean refractive index $\langle \bar{n}(r)\rangle$ around each dielectric junction.  In addition, the average refractive index for both networks approaches the same value at large $r$ since the air fraction $\gamma$ is the same for both structures.  However, the variance $V(r)$ of $\bar{n}$ for the two network structures shows marked differences for all $r$ as shown in Fig.~\ref{fig4}(b).  The tetrahedral network possesses much smaller fluctuations in $\bar{n}$ from one junction to another.  Thus, the scattering units are much more uniform for the tetrahedral network than those in the air spheres.  The uniformity of local refractive index distribution ensures similar scattering characteristic of individual dielectric junctions and facilitates their coupling which leads to a dramatic depletion of the DOS.\\

\begin{figure}[htbp]
\centering
\includegraphics[scale = 0.5]{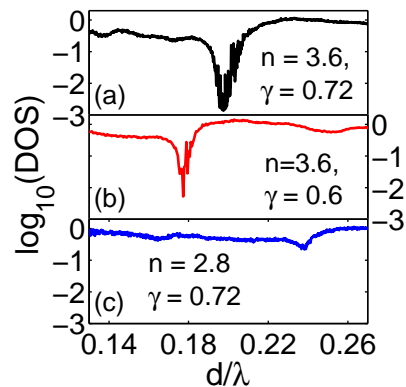}
\caption{DOS of tetrahedral networks for two values of the air fraction $\gamma$ and refractive index $n$. (a) $n = 3.6$ and $\gamma = 0.72$, (b) $n = 3.6$ and $\gamma = 0.6$, and (c) $n = 2.8$ and $\gamma = 0.72$.    }
\label{fig5}
\end{figure}

\indent The formation of a PBG in the tetrahedral network structure also depends on the air fraction $\gamma$ and the refractive index of the dielectric material $n$.  In Fig.~\ref{fig5}, we show the variation of the PBG for two values of $\gamma$ and $n$.  Reducing the air fraction below $0.8$ leads to a decrease in the PBG.  A reduction in $\gamma$ increases the average refractive index of the structure, thus reducing the ratio of the index difference $(n-1)$ to the average refractive index.  It leads to a decrease of the overall scattering strength.  In contrast, if $\gamma$ is increased above $0.8$, there is an insufficient amount of high-index material to scatter light.  Thus, an optimal $\gamma$ that depends on the refractive index $n$ exists at which the scattering strength is maximal and the PBG is the largest.  As shown in Fig.~\ref{fig5} (c), when the refractive index of the dielectric material is reduced from $n=3.6$ to $n=2.8$, the maximal DOS reduction shifts to $\gamma = 0.72$.  Moreover, the reduction of the DOS at $n=2.8$ is nearly two orders of magnitude smaller than that at $n=3.6$.  Thus, a large refractive index contrast is needed for strong PBGs in photonic amorphous structures. \\

\section{Effect of short-range order}
\label{order}
\begin{figure}[htbp]
\centering
\includegraphics[scale = 0.28]{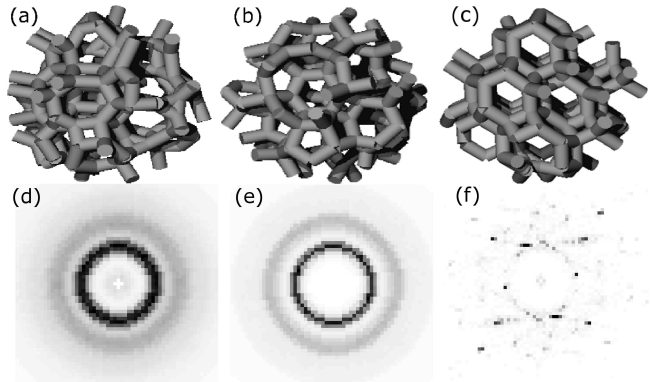}
\caption{Tetrahedral dielectric networks generated from sphere packings with packing fraction (a) $\phi = 0.35$, (b) $0.64$, and (c) $0.69$. 2D cross-sections of the 3D spatial Fourier spectra of the corresponding tetrahedral networks are shown in (d), (e), and (f).}
\label{fig6}
\end{figure}

\indent In addition to the factors studied above, short-range positional order and tetrahedral bond order play  important roles in the formation of PBGs in PAS.  
In this section, we focus on the dielectric network of tetrahedral bonding, which yields the largest PBGs, and vary the amount of positional and tetrahedral bond order.  
In particular, we tune the positional order of the original sphere packings from which the tetrahedral networks are formed.  The degree of positional order increases with the volume fraction of spheres $\phi$, which varies from $0.35$ to $0.69$.
  We label the tetrahedral networks (Fig.~\ref{fig6}(a)-(c)) generated from the sphere packings at $\phi= 0.35$, $0.64$, and $0.69$ as $A$, $B$, and $C$.  2D cross-sections of the 3D spatial Fourier spectra for these structures are presented in Fig.~\ref{fig6} (d)-(f).  The power spectra of networks $A$ and $B$ consist of concentric shells, but the shell width is notably larger for $A$.  Thus both $A$ and $B$ are isotropic structures, but $B$ possesses more positional order than $A$.  In contrast to $A$ and $B$, network $C$ features discrete diffraction peaks in the Fourier spectrum, and the structure is no longer isotropic.\\

\begin{figure}[htbp]
\centering
\includegraphics[scale = 0.5]{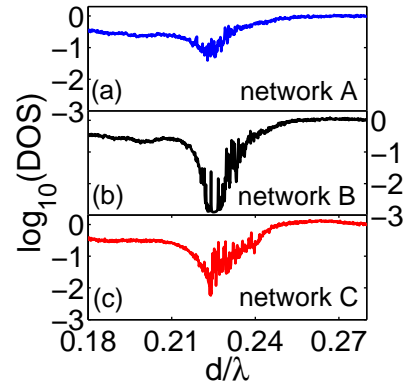}
\caption{The DOS for three tetrahedral dielectric networks (a) $A$, (b) $B$, and (c) $C$ with positional order increasing from $A$ to $C$. }
\label{fig7}
\end{figure}

\indent In Fig.~\ref{fig7}, we compare the DOS of the tetrahedral networks $A$, $B$, and $C$, with the refractive index of the dielectric rods set to $n=3.6$.  By adjusting the dielectric rod radius $W$, we find that the optimal air fraction for all three structures is $\gamma = 0.8$.  As expected, network $A$, with the least positional order, possesses the smallest depletion in the DOS.  However, network $C$ with the strongest degree of positional order has a smaller DOS depletion than network $B$.  This result contrasts with recent findings for 2D PAS with air cylinders embedded in dielectric materials that show increasing positional order leads to stronger DOS depletion~\cite{jin}.  To understand these results, we must also compare the uniformity of the local refractive index distribution and the structural topology of the three network structures at fixed radius $W$ of the dielectric rods.  We find that networks $B$ and $C$ have comparable fluctuations in $\bar{n}$ over all the junctions. Thus, local uniformity does not explain the difference in the depletion of the DOS for networks $B$ and $C$.\\ 

\begin{figure}[htbp]
\centering
\includegraphics[scale = 0.46]{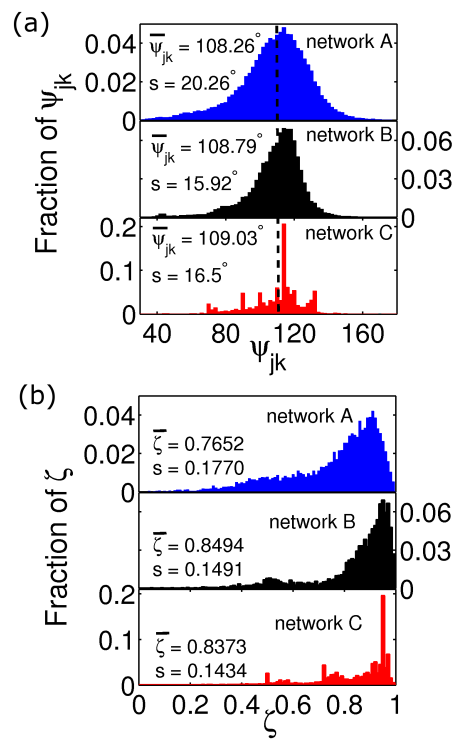}
\caption{Characterization of the local topology for networks $A$, $B$ and $C$. (a) The distribution of angles $\psi_{jk}$ between dielectric rods $j$ and $k$ at each tetrahedral junction. The vertical dashed line indicates the angle for the periodic diamond structure, $\psi_{jk} = 109.5^\circ$. (b) Distribution of the tetrahedral order parameters $\zeta$ at each junction.  The average ${\bar \psi}_{jk}$ and ${\bar \zeta}$ and standard deviations $s_{\psi}$ and $s_{\zeta}$ are also provided.}
\label{fig8}
\end{figure}

\indent To investigate the effects of local topology on the depletion of the DOS, we compute the tetrahedral order parameter~\cite{steinhardt,notomi2} 
\begin{eqnarray}
\zeta &=& 1 - \frac{3}{8}\sum_{j=1}^{3}\sum_{k=j+1}^{4}\left( \cos \psi_{jk} +\frac{1}{3}\right)^2, 
\end{eqnarray}  
where $\psi_{jk}$ is the angle between two dielectric rods joined at a junction in the tetrahedral network~\cite{hardwick}.  For a periodic diamond network, $\psi_{jk}=109.5^\circ$, $\cos(\psi_{jk}) = -1/3$ for all $j$ and $k$, and thus $\zeta = 1$ at each junction.  If the dielectric rods are randomly orientated, $\langle \zeta \rangle = 0$.  In Fig.~\ref{fig8}, we plot the distributions of $\psi_{jk}$ and $\zeta$ for the $A$, $B$, and $C$  networks, and provide the mean values ($\bar{\psi}_{jk}$ or $\bar{\zeta}$), and standard deviations $s_{\psi}$ and $s_{\zeta}$.\\

\indent Network $A$ possesses the widest distributions for both $\psi_{jk}$ and $\zeta$, which indicates that the local topology varies significantly from one junction to another and the bond angles within each junction are not uniform.  The distributions of $\psi_{jk}$ and $\zeta$ are narrower for network B, and are peaked at $\psi_{jk} = 114^\circ$ and $\zeta = 0.95$, which indicates that most of the junctions have a similar topology to that in a diamond lattice.  In contrast, network $C$ displays multi-modal distributions for $\psi_{jk}$ and $\zeta$.  For example, the $\zeta$ distribution possesses peaks at $\zeta = 0.95$, $0.72$, and $0.5$.  The first peak reveals that there are many junctions with strong tetrahedral order, while the second and third peaks reflect the existence of many ``defect'' junctions with low $\zeta$.  Such defect junctions are likely located at domain boundaries, and introduce irregularity in the local configuration of scattering units.  Figures~\ref{fig7} and~\ref{fig8} show that photonic amorphous networks with strong tetrahedral order and few defect junctions have broad PBGs. \\

\section{Conclusion}\label{last}  
In this paper, we calculate the DOS in 3D photonic amorphous structures with cermet and network topologies.  We find that interconnected networks of high-index material with uniform dielectric junctions and tetrahedral bonding give rise to large isotropic PBGs.  Further, reduced fluctuations in the refractive index around each junction and strong tetrahedral order for the angles between the dielectric rods that form the junctions enhance isotropic PBGs.  High refractive index contrast and a low fraction of high-index material are also important to PBG formation.  We have thus identified several parameters that can be tuned to create broad isotropic PBGs in photonic amorphous structures in the absence of long-range structural order. \\

\section{Acknowledgments}
\indent We thank W. L. Vos and R. O. Prum for useful discussions.
This work was supported with seed funding from the Yale NSF-MRSEC (DMR-0520495) and NSF grants to HC (PHY-0957680), ERD (CAREER CBET-0547294) and CSO (DMS-0835742). J.-K. Yang acknowledges support from the National Research Foundation of Korea under grant No. NRF-2009-352-C00039.  
This work also benefited from the facilities and staff of the Yale University Faculty of Arts and Sciences High Performance Computing Center and NSF grant No. CNS-0821132 that partially funded acquisition of the computational facilities.


\end{document}